\documentclass[%
preprintnumbers,
 amsmath,amssymb,
 aps,
]{revtex4-2}

\usepackage{graphicx}
\usepackage{dcolumn}
\usepackage{bm}

\begin{document}

\title{Quantum-limited measurements of intensity noise levels in Yb-doped fiber amplifiers}

\author{Alexandra Popp$^{1,2,3}$} \email{alexandra.popp@mpl.mpg.de} \author{Victor Distler$^{4,5}$} \author{Kevin Jaksch$^{1,2,3}$} \author{Florian Sedlmeir$^{1,2}$} \author{Christian R. M\"uller$^{1,2}$} \author{Nicoletta Haarlammert$^{4}$} \author{Thomas Schreiber$^{4}$} \author{Christoph Marquardt$^{1,2}$} \author{Andreas T\"unnermann$^{4,5}$} \author{Gerd Leuchs$^{1,2}$}
 
\address{$^{1}$Max$\,\,$Planck$\,\,$Institute$\,\,$for$\,\,$the$\,\,$Science$\,\,$of$\,\,$Light,$\,\,$Staudtsta\ss e$\,\,$2, 91058$\,\,$Erlangen,$\,\,$Germany\\
$^{2}$Institute$\,\,$of$\,\,$Optics,$\,\,$Information$\,\,$and$\,\,$Photonics,$\,\,$Friedrich-Alexander$\,\,$University$\,\,$Erlangen-N\"urnberg,$\,\,$Staudtstra\ss e$\,\,$7$\,\,$B2,$\,\,$91058$\,\,$Erlangen,$\,\,$Germany\\
$^{3}$Erlangen$\,\,$Graduate$\,\,$School$\,\,$in$\,\,$Advanced$\,\,$Optical$\,\,$Technologies$\,\,$(SAOT),$\,\,$Friedrich-Alexander$\,\,$University$\,\,$Erlangen-N\"urnberg,$\,\,$Paul-Gordan-Stra\ss e$\,\,$6,$\,\,$91052$\,\,$Erlangen,$\,\,$Germany\\
$^{4}$Fraunhofer$\,\,$Institute$\,\,$for$\,\,$Applied Optics$\,\,$and$\,\,$Precision$\,\,$Engineering,$\,\,$Albert-Einstein-Stra\ss e$\,\,$7,$\,\,$07745$\,\,$Jena,$\,\,$Germany\\
$^{5}$Friedrich-Schiller-University$\,\,$Jena,$\,\,$Institute$\,\,$of$\,\,$Applied Physics,$\,\,$Albert-Einstein-Stra\ss e$\,\,$15,$\,\,$07745$\,\,$Jena,$\,\,$Germany}

\begin{abstract}
We investigate the frequency-resolved intensity noise spectrum of an Yb-doped fiber amplifier down to the fundamental limit of quantum noise. We focus on the kHz and low MHz frequency regime with special interest in the region between 1 and 10$\,$kHz. Intensity noise levels up to $\ge$60$\,$dB above the shot noise limit are found, revealing great optimization potential. Additionally, two seed lasers with different noise characteristics were amplified, showing that the seed source has a significant impact and should be considered in the design of high power fiber amplifiers.
\end{abstract}

\maketitle

\section{Introduction}

High power fiber amplifiers have become a widely used tool in research and industry \cite{jauregui} due to their robust operation, excellent beam quality and power scalability. Some applications directly pose high requirements on the stability and the intensity noise characteristics of fiber amplifiers, such as their potential usage as laser sources in next-generation gravitational wave detectors  \cite{kracht}. Other applications, such as laser cutting, require low noise more indirectly, since single mode power scalability is currently limited by transverse mode instabilities (TMI) \cite{zervas2}. The onset of these TMI is in turn influenced by the intensity noise in such fiber lasers \cite{stihler}. These insights have recently put the  noise characteristics of fiber amplifiers into the focus of several investigations \cite{gierschke, li, zhao, buikema}. In previous studies the relative intensity noise was used as a figure of merit, whereas the information on the absolute noise level with respect to the theoretical boundary remains unknown. \newline
\indent The theoretically possible minimum noise in any laser is defined by the quadrature variance of the coherent state \cite{clerk}, the so called shot noise limit (SNL). It is known that phase insensitive amplification processes add noise to the input state and therefore degrade the noise performance of the system \cite{haus, caves}. This is the well known 3$\,$dB limit of optical amplification \cite{bachor}. While quantum noise as well as its impact have been investigated in various phase sensitive and insensitive amplifier systems bevore \cite{kuo, levenson}, it has not yet been experimentally studied in detail for Yb-doped high power fiber amplifiers. An approach to improve the noise characteristics of amplifier systems could be coherent combining of multiple amplifiers. In this case multiple beams are amplified separately and then recombined to generate a high power output \cite{liuCC}. Recently, the noise performance of these systems has been theoretically investigated down to the quantum mechanical noise limit for the first time, showing that it can outperform the output noise variances of a linear amplifier under certain conditions \cite{mueller}. A different attempt for optimizing the noise properties of amplifier systems is to reduce the noise level of its input state - the seed laser - as far as possible. A precise knowledge of the amplifier noise with respect to the SNL might allow to assess the future potential of high power fiber amplifiers for noise sensitive fields of research like high precision optical frequency comb generation and quantum metrology \cite{udem, brandt}.\newline
\indent High power fiber systems usually consist of several amplification stages \cite{beier, guiraud, zhao2}. In this contribution, we examine the first amplification stage of such an Yb-doped fiber amplifier chain. The noise transfer functions of rear amplification stages have been measured, and it has been shown that they can dampen the relative intensity noise \cite{gierschke}. However, the total amount of noise in such systems remained unclear. Thus, this contribution is focused on the amplifier parts, that are expected to have the highest noise, the laser seed and the first amplification stage, and presents  intensity noise measurements of such a system relative to the shot noise limit. This work is organized in the following manner: In section 2 an overview of the experimental setup is given. Section 3 presents and discusses the results, which are brought to a conclusion in section$\,$4.

\section{Experimental setup}

\begin{figure}[h!]
\centering\includegraphics[width=10cm]{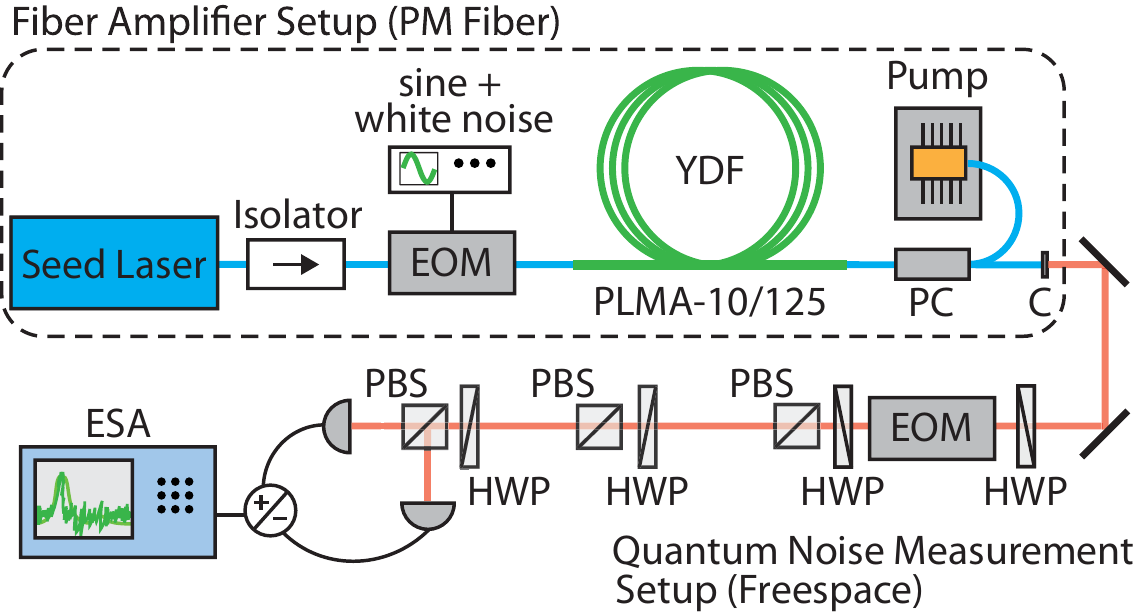}
\caption{Schematic diagram of the Yb-doped fiber amplifier with quantum noise measurement setup. PM Fiber: Polarization maintaining fiber, EOM: electro-optic modulator, YDF: Ytterbium doped fiber, PC: pump combiner, C: collimator, HWP: half-wave plate, PBS: polarizing beamsplitter, ESA: electronic spectrum analyzer.}
\end{figure}

As shown in Figure 1, the experimental setup consisted of two parts: the examined laser source and the quantum noise measurement setup. The laser seed source was an external cavity diode laser at 1064 nm that was phase-modulated to avoid Stimulated Brillouin Scattering in subsequent parts of the laser \cite{harish}. After phase modulation, the light, with a power of 5 mW, was amplified in a PLMA-10/125 Ytterbium-doped fiber to an output power of 1.89 W, corresponding to a gain of approximately $26\,$dB. The pump light at a wavelength of 976 nm was coupled in counter-propagating direction by a pump combiner. Fiber connectors after the seed laser and the EOM allowed to examine these parts individually. At the end of this all-fiber polarization maintaining setup, a collimator was used to couple the light to the measurement setup. \newline
\indent The latter consisted of a typical self-homodyne measurement setup \cite{bachor} and an electro-optic modulator used for checking the common mode rejection capability of the system. A half-waveplate (HWP) and polarizing beamsplitter (PBS) after the modulator were used to transform the created polarization modulation to amplitude modulation. Another set of HWP and PBS were deployed to be able to attenuate the optical power during the measurements. A final HWP and PBS split the signal. The splitted signals were detected with two identically constructed, home built detectors. The signals of the two detectors were added or subtracted by in-house produced electronics. In the difference signal, the classical noise contribution is canceled out and only quantum noise of the light field and electronic noise of the detectors are visible. In the sum signal, the quantum as well as the classical intensity noise of both signals are measured \cite{bachor}. Measurements were acquired either with an electronic spectrum analyser (ESA) or alternatively using an oscilloscope for the low frequency regime. To align the last set of HWP and PBS, a sinusoidal signal was modulated onto the laser using the EOM and monitored the difference signal with the ESA. By adjusting the waveplate to equalize the power in the two arms, we were able to suppress the amplitude modulation with a typical common mode rejection of $\ge$30$\,$dB. 

\section{Experimental results and discussion}

In order to investigate the intensity noise of our Yb-doped fiber amplifier we characterized the three main components (seed laser, phase modulator and amplifier) separately for different frequency regimes ranging from the low kHz to the MHz domain. Furthermore, we investigated two different seed laser sources, to gain more insight on the contribution of the seed laser intensity noise in fiber amplifiers. Moreover, we relate these noise measurements to the carefully calibrated shot noise limit.

\begin{figure}[]
\centering\includegraphics[width=12cm]{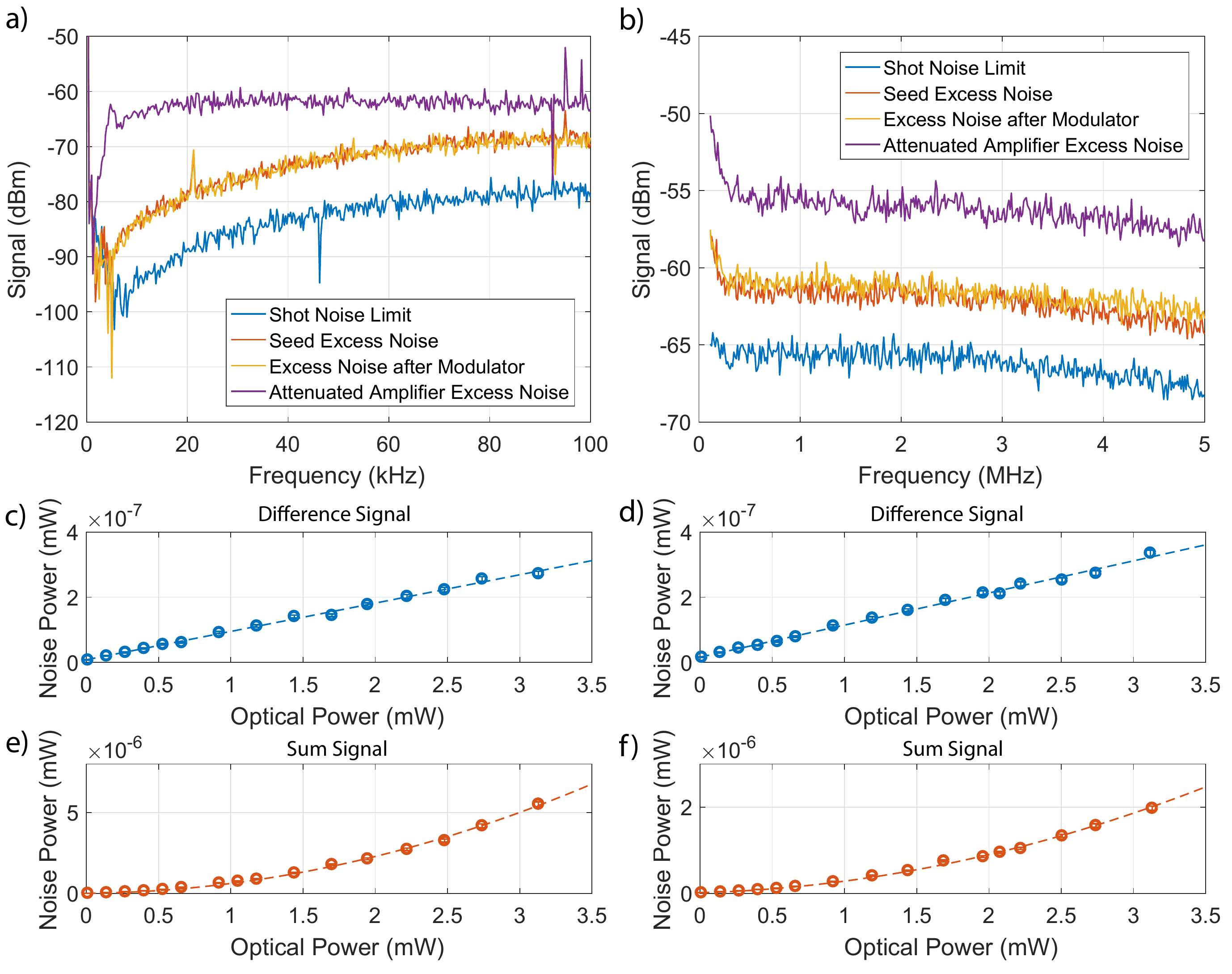}
\caption{Laser and amplifier amplitude noise spectrum with respect to shot noise a) in the 0-100kHz regime with RBW and VBW of 10Hz b) in the 0.1-5 MHz regime with RBW and VBW of 1kHz. Figures c) and d) show the linear noise over optical power dependency of the difference signal (shot noise) for 50kHz and 3MHz, respectively. Figures e) and f) contain the respective quadratic excess noise dependencies of the sum signal.}
\end{figure}

\indent Figure 2a) and b) show frequency resolved intensity noise spectra for the frequency regimes from 0-100$\,$kHz (a) and 0.1-5$\,$MHz (b). The strong, frequency dependent noise increase in the range between 0-100$\,$kHz is caused by the frequency dependent gain of our ESA. Therefore all noise contributions should be considered by their relative value above the SNL. The SNL was measured by subtracting the signals of the two individual detectors, while the excess noise is given by the added signals as discussed in section 2. To confirm the quantum origin of the measured SNL, we attenuated both the sum and the difference signal over a range of 15$\,$dB in optical power. The corresponding results are presented in Figures 2c) - f). To generate the attenuation measurements, zero-span measurements containing 400 datapoints at a given frequency (50$\,$kHz and 3$\,$MHz, as shown in Figures 2c)-f)) with a bandwidth of 1$\,$kHz were used to obtain reliable amplitude noise values for a given optical power. Figures 2c) and d) hint a clear linear dependence of the noise on optical power for the difference signal. In contrast, the quadratic noise to optical power dependency in Figures 2e) and f) is created by the sum signal. This provides us with strong evidence for the quantum and classical nature of the respective signals \cite{walls}.  Returning to Figures 2a) and b), the seed laser clearly showed excess noise above the shot noise limit (SNL) of about 9$\,$dB in the kHz domain, decreasing to about 4$\,$dB in the MHz domain.  Next we introduced the SBS preventing phase modulator to our system. It is evident that it did not generate additional intensity noise, as expected when properly aligned. Finally, the amplifier was introduced. Here, dispersion could transfer the phase noise to intensity noise, but this effect is, given the narrow laser linewidth ($140\,$pm), short fiber length ($4\,$m) and low modulation frequency (maximal $525\,$MHz), negligible. The amplifier created approximately 26$\,$dB of amplification in terms of optical power. Therefore the amplifier was attenuated by 26$\,$dB, in order to be compatible with the power requirements of the quantum noise measurement setup. The noise level of the amplifier was significantly higher than of the seed laser, indicating that the amplification process was a major noise contribution. Following \cite{tunnermann}, the relative intensity noise transfer function of the pump is a low-pass, which explains the observed frequency dependency of the noise gain shown in Figure 2a). At 10 kHz, we measured the attenuated amplifier to have approximately 34$\,$dB of intensity noise above the SNL. In the MHz regime, the amplifier excess noise follows the seed noise spectrum introducing about 6$\,$dB of intensity noise after attenuation. For the noise contribution at full amplifier power, the introduced optical attenuation as well as all losses have to be considered. To get a best estimate of the excess noise above SNL for all frequencies, we measured the noise spectra given in Figure 2a) and b) for range of optical powers and fitted the attenuation behavior independently for the four cases of SNL, seed excess noise, modulator excess noise and the attenuated amplifier excess noise to extract the exact values of excess noise above SNL for the latter. The excess noise  for the unattenuated amplifier was calculated by inserting the full optical power into the fit of excess noise above SNL. Exemplary values for the full amplifier excess noise are given in Table 1.

\begin{table}[]
\centering
\begin{tabular}{c c c}
& Seedlaser Excess Noise & Amplifier Excess Noise \\ 
Frequency &  ($P_\text{opt}$ = 1.3$\,$mW) (above SNL) &  ($P_\text{opt}$ = 1.89$\,$W) (above SNL) \\
\hline
\hline
10$\,$kHz &  11.9$\,$dB & 62.8$\,$dB \\ 
50$\,$kHz &  7.8$\,$dB & 50.6$\,$dB \\ 
90$\,$kHz &  8.2$\,$dB & 49.3$\,$dB \\ 
1$\,$MHz &  4.5$\,$dB & 42.1$\,$dB \\ 
4$\,$MHz &  4.2$\,$dB & 41.6$\,$dB \\
\hline
\hline
\end{tabular}
\caption{Exemplary values for unattenuated amplifier excess noise given in dB above SNL}
\end{table}
To interpret the values given in Table 1, we need to consider the theoretical noise limit of the amplification process. This would be bounded by using a shot-noise limited seed laser and a perfect, quantum limited linear amplifier. For this case, we expect 29$\,$dB of noise in our system for an optical gain of 26$\,$dB \cite{bachor}. This means, that we exceed the theoretical limit for the linear amplifier by at least a factor of 12$\,$dB in the MHz regime, at 10$\,$kHz we transcend the theoretical limit by more than 33$\,$dB. \\
\indent As previously discussed in section 1, the current power scaling limit in Yb-doped fiber amplifiers is set by transverse mode instabilities \cite{zervas2}. Typically, our high power amplifiers exhibit TMI with frequencies up to 10$\,$kHz \cite{beier}, thus this frequency regime is especially interesting in our analysis. To investigate this domain in more detail, we exchanged the electronic spectrum analyzer (ESA) with an oscilloscope. The recorded time traces of 1$\,$s with a sampling rate of 1$\,$Ms/s were Fourier transformed to obtain the magnitude of the signals' power spectral density (PSD) as shown in Figure 3a). Using two input ports of the oscilloscope allowed us to monitor the sum and difference signals simultaneously. In agreement with the previous ESA measurements, we found a flat seed excess noise spectrum with approximately 10$\,$dB above SNL and no additional contribution from the phase modulator. The large variance of the traces can most likely be attributed to the theoretical resolution bandwidth of 1$\,$Hz and could be potentially smoothed out by introducing an artificial larger bandwidth into the measurement. Similar to the previous measurements, we attenuated the optical power over a range of approximately 10$\,$dB. We summed 100 consecutive points of the trace allowing us a measurement with a bandwidth of 0.1$\,$kHz. Exemplary scaling curves are shown in Figure 3b) and c), again confirming the quantum nature of the difference signal in contrast to the excess noise. The attenuated amplifier noise spectrum is mainly flat at a level of 30 dB above SNL. However, there are two distinct features in the spectrum: a broad peak of about 10 dB between 4 to 5 kHz and a higher peak in the regime below 0.5 kHz. The latter is also visible in the SNL trace and thus most likely attributed to electronic circuit noise of the detectors. The broader peak on the other hand stems from the amplification process but its precise origin remains unclear. Potentially, the pump diodes driving electronics cause this contribution. Following the above procedure to account for the optical attenuation, we get a total amplifier excess noise of 67.5$\,$dB above SNL at 3$\,$kHz, again by far exceeding the theoretical limit of 29$\,$dB. \\

\begin{figure}[]
\centering\includegraphics[width=12cm]{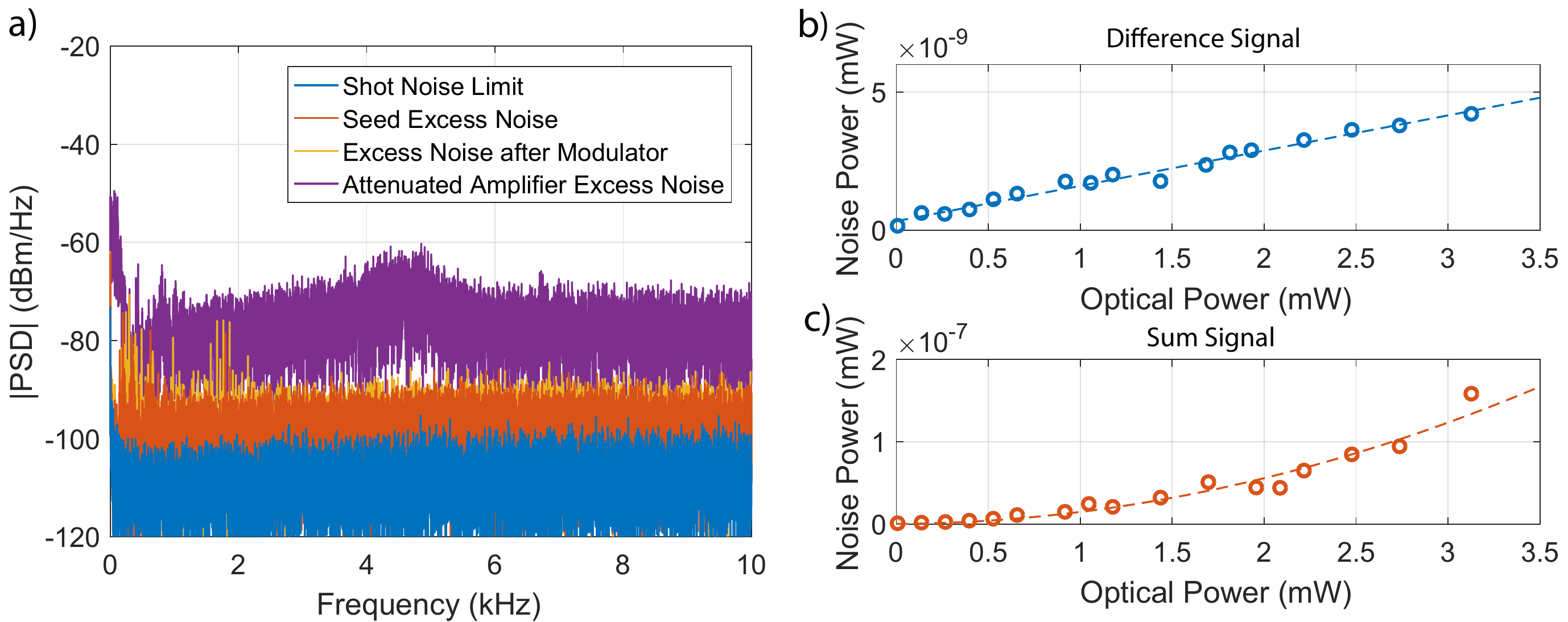}
\caption{Measurement of low frequency intensity noise power spectral density in the regime between 0-10$\,$kHz (a) and respective scaling curves to show the noise over optical power dependency for the difference (shot noise) (b) and sum (c) signal at 3$\,$kHz.}
\end{figure}
\begin{figure}[]
\centering\includegraphics[width=12cm]{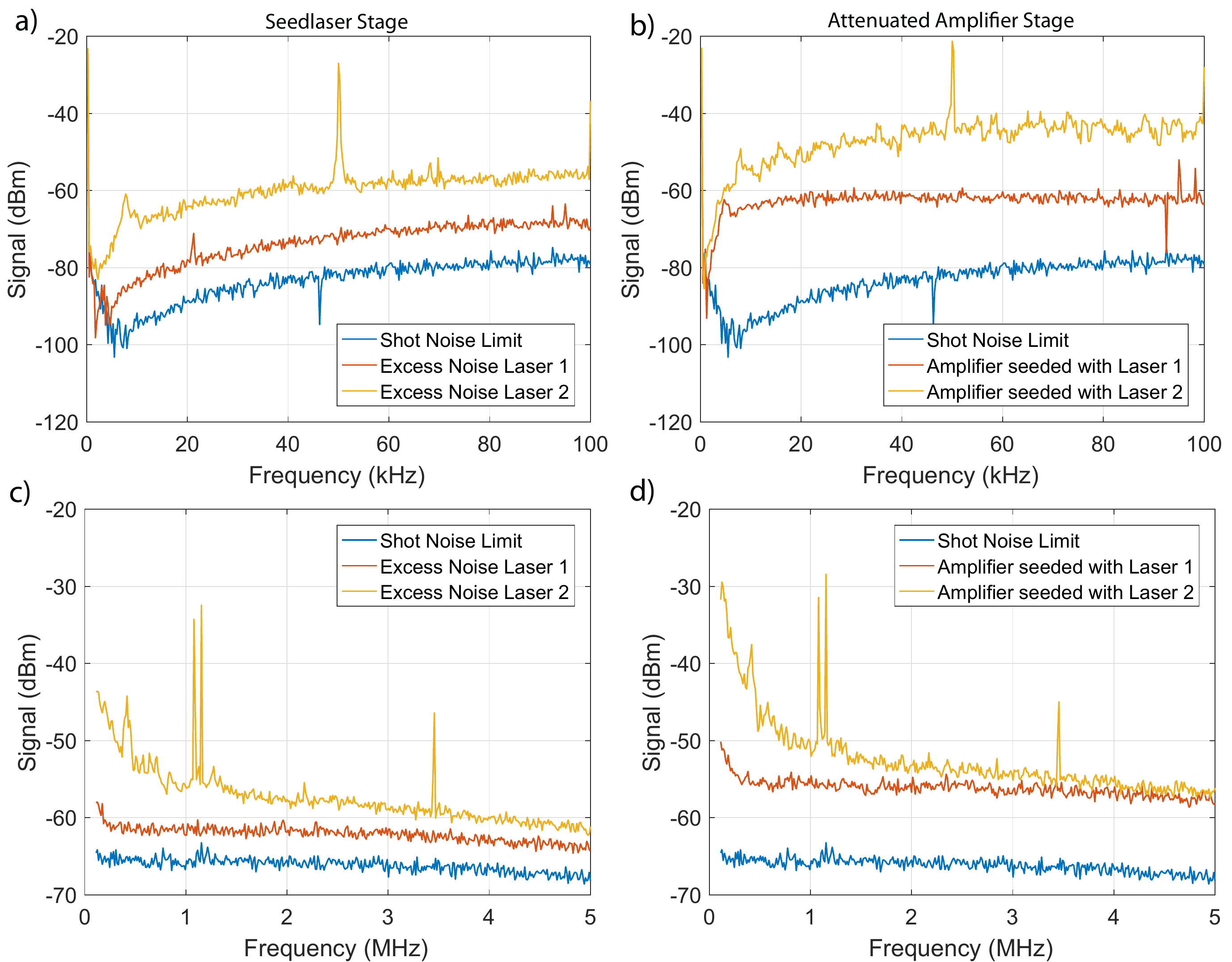}
\caption{The diagrams show the comparison of two different seedlasers (laser 1 and laser 2) before (a, c) and after (b, d) propagation through the fiber amplifier in the kHz (a, b) and MHz (c, d) frequency range.}
\end{figure}

\indent As a final step, we exchanged our seed laser to a comparable ECDL with larger amplitude noise, denoted as laser$\,$2, and repeated our previous measurements. The complete spectral overview for the seed laser as well as attenuated amplifier stage is given in Figure 4. Consistent with previous measurements from this section, the phase modulator does not introduce additional amplitude noise and is thus excluded in further analysis. Figure 4a) and c) present the two different seed lasers for the kHz (a) and MHz (c) domain. It is evident that the noise level of laser$\,$2 exceeded the one of laser$\,$1 by more than 10$\,$dB in the kHz regime. The excess noise of laser 2 decreased for larger frequencies ending up with approximately 2$\,$dB above the level of laser$\,$1 at 5$\,$MHz. While the amplifier conserved the observed spectrum of both lasers in the MHz measurement (Figure 4d)), the noise properties in the kHz regime are vastly different. The noise increase towards low frequencies is less pronounced in the amplifier seeded with laser$\,$2, while at the same time the variance of the noise trace was heavily enhanced. A possible explanation for this behavior would be given by the large enhancement of the seednoise contribution, potentially exceeding the low frequency pump noise dominance given in the amplifier seeded with laser$\,$1. This result demonstrates, that the seed laser intensity noise characteristics can have great influence on the excess noise of the fiber amplifier in the kHz as well as in the low MHz frequency regime.
%
\section{Conclusion}

We characterized the intensity noise of an Yb-doped fiber amplifier with respect to shot noise for a variety of different frequency regimes. We identified a significant noise contribution above the shot-noise limit (SNL) in the kHz as well as in the MHz regime. Special interest was given to the region below 10$\,$kHz, where we found the highest intensity noise contribution ($\ge$60$\,$dB above SNL) in our investigation. Furthermore we showed that in our case, the choice of seed laser and its intensity noise had a significant impact on the final noise performance of the amplifier. It can be concluded, that commonly used amplifiers still have a great potential of noise reduction.

\section*{Funding}
Fraunhofer and Max Planck cooperation program (PowerQuant)

\section*{Acknowledgments}
The authors acknowledge the support of Oliver Bittel and Lothar Meier.

\bibliography{References}

\end{document}